\documentclass[letterpaper,superscriptaddress,english,aps,prl,twocolumn,floatfix, showpacs, amsfonts, amssymb]{revtex4}
\usepackage{epsfig, graphicx,psfrag,amsmath,amssymb,float}
\usepackage[T1]{fontenc}
\usepackage[latin9]{inputenc}
\usepackage{amsmath}
\usepackage{amssymb}
\usepackage{bbold}
\usepackage{amscd}
\usepackage{bm}
\usepackage{psfrag}

\usepackage{babel}

\begin{document}
\title{Dynamical structure factors of the spin-$1/2$ $XXX$ chain at finite magnetic field}

\author{J. M. P. Carmelo}
\affiliation{Department of Physics, University of Minho, Campus Gualtar, P-4710-057 Braga, Portugal}
\affiliation{Center of Physics of University of Minho and University of Porto, P-4169-007 Oporto, Portugal}
\affiliation{Beijing Computational Science Research Center, Beijing 100084, China}
\author{P. D. Sacramento}
\affiliation{CeFEMA, Instituto Superior T\'ecnico, Universidade de Lisboa, Av. Rovisco Pais, P-1049-001 Lisboa, Portugal}
\author{D. K. Campbell}
\affiliation{Boston University, Department of Physics, Boston, MA 02215 USA}

\date{20 March 2015}

\begin{abstract}
We study the dynamical structure factors of the spin-$1/2$ $XXX$ chain at finite magnetic field $h$, focusing in particular
on the singularities at excitation energies in the vicinity of the lower thresholds $\omega_{L}^{\tau} (k)$ of the 
leading-order dominant excitations. We derive the exact momentum and spin-density dependences of the exponents 
$\zeta^{\tau} (k)$ controlling the singularities for both the longitudinal $(\tau = l)$ and transversal $(\tau = t)$ dynamical 
structure factors for the whole momentum range $k\in [0,\pi]$, in the thermodynamic limit. In that limit we argue that 
the higher-order excitations change neither $\zeta^{\tau} (k)$ nor the sharpness of the singularities. We discuss the 
relation to neutron scattering and suggest new experiments on spin-chain compounds using a carefully oriented crystal.
\end{abstract}


\maketitle

The anisotropic spin-$1/2$ Heisenberg $XXZ$ chain in a magnetic field $h$ with anisotropy parameter $\Delta\ge 0$ and exchange 
integral $J$ is a paradigmatic example of an integrable strongly correlated system \cite{Taka-AN}. Its Hamiltonian is,
\begin{equation}
\hat{H} = J\sum_{j=1}^{L}\left(\sum_{\tau =x,y}{\hat{S}}_j^{\tau}{\hat{S}}_{j+1}^{\tau} + \Delta\,{\hat{S}}_j^z{\hat{S}}_{j+1}^z\right) 
+ g\mu_B\,h\sum_{j=1}^{L} \hat{S}_j^z \, ,
\label{H}
\end{equation}
where for simplicity we take $g=2$, $\mu_B$ is the Bohr magneton, and $\hat{S}_j^{x,y,z}$ are components of the 
spin-$1/2$ operators at site $j=1,...,L$. We consider states with spin $S\in]0,L/2]$ and spin density $m = 2S^z/L \in ]0,1]$
where $S^z={1\over 2}(L-2M)$ is the spin projection and $M$ the number of down spins.

At the isotropic point, $\Delta=1$, the $XXZ$ chain becomes the $XXX$ chain \cite{Taka,Muller} and contains antiferromagnetic 
correlations that have been observed in dynamical quantities measured in experiments on spin-chain compounds
\cite{SPA-11,Stone-03,Thurber-01,Motoyama-96}. For instance, its dynamical structure factors,
\begin{equation}
S^{\alpha\alpha} (k,\omega) = \sum_{j=1}^L e^{-ik j}\int_{-\infty}^{\infty}dt\,e^{-i\omega t}\langle 0\vert\hat{S}^{\alpha}_j (t)\hat{S}^{\alpha}_j (0)\vert 0\rangle \, ,
\label{SDSF}
\end{equation}
where $\alpha =x,y,z$ are important examples of relevant quantities from the experimental viewpoint \cite{Stone-03}.

The isotropic $\Delta=1$ model also poses the most challenging technical 
problems for theory. This holds true also for $h=0$, which is the case that has been most heavily investigated 
\cite{Bougourzi-97,Caux-06,Imambekov-09,Caux-11}. Indeed, as shown in Ref. \cite{Caux-11}, for $\Delta \in [0,0.8]$ and $h=0$
nearly all the $S^{zz} (k,\omega)=S^{xx} (k,\omega)$ integrated intensity spectral weight stems from two-hole Bethe ansatz (BA) excitations 
whereas as $\Delta$ increases from $0.8$ to $1.0$, the four-hole excitations increasingly contribute 
as $\Delta\rightarrow 1$. Specifically, in that limit the contributions to $S^{zz} (k,\omega)$ 
from the $S=1$ and $S^z=0$ two-hole excitations correspond to a relative integrated intensity of $\approx 0.75$ \cite{Bougourzi-97}.
This increases to $\approx 0.99$ if one considers in addition $S=1$ and $S^z=0$ four-hole excitations. Importantly, however, 
all the $S^{zz} (k,\omega)$ singularities are determined by contributions 
from two-hole excitations and the exponent that controls them can be shown from purely phenomenological 
considerations to be fixed at $h=0$ by the $SU(2)$ invariance alone \cite{Imambekov-09}. 

At $h> 0$ one has that $S^{zz} (k,\omega)\neq S^{xx} (k,\omega)=S^{yy} (k,\omega)$. Previous studies on these 
dynamical structure factors in the $XXX$ model refer to {\it finite systems} and rely on numerical diagonalizations \cite{Lefmann-96}, 
evaluation of matrix elements between BA states \cite{Muller,Karbach-00,Karbach-02}, and the form-factor method \cite{Biegel-02},
which applies to the $XXZ$ chain \cite{Kitanine-99,Caux-05}. The studies of Refs. \cite{Caux-05,Rodrigo-08} refer mostly to $\Delta < 1$. 
The line shape in the vicinity of the dynamical structure factor thresholds is predicted to be controlled by 
momentum dependent exponents \cite{Rodrigo-08}, but that dependence remains an important open problem.

On the other hand, our present study and results refer to the {\it thermodynamic limit} (TL). Specifically, we clarify one of the unsettled questions concerning 
the physics of the spin-$1/2$ $XXX$ Heisenberg model at finite magnetic field $h$ by deriving the exact momentum and spin-density 
dependences of the exponents that control the lower thresholds singularities in $S^{zz} (k,\omega)$
and $S^{xx} (k,\omega)$. We discuss the relation of our theoretical results to the $\omega$ dependence of the magnetic 
scattering intensity observed in inelastic neutron scattering \cite{Stone-03} and suggest new experiments 
using a carefully oriented crystal.

Our calculation approach uses the pseudofermion dynamical theory (PDT) developed in Refs. \cite{V-1,LE} for the one-dimensional 
(1D) Hubbard model in the TL \cite{Hubbard-94}, which can also be used for the $XXX$ chain. When the lower thresholds 
of the spectral and dynamical functions refer to the branch lines as defined within that theory, the exact line shape in the 
vicinity of such thresholds is indeed controlled by momentum-, density-, and interaction-dependent exponents. These are 
expressed in terms of dressed phase shifts that are known from the BA solution. 

After the PDT was introduced, a set of novel methods were developed to tackle also the finite-energy physics
of integrable and non-integrable 1D correlated quantum problems, beyond the low-energy limit \cite{Essler-14,Glazman-12,DSF-n1}. 
For instance, the same exact momentum dependence derived by the PDT for the exponents that 
control threshold singularities of the 1D Hubbard model one-electron spectral functions \cite{V-1} has recently 
been obtained in the framework of a mobile impurity approach \cite{Essler-14}. 

For $h> 0$ the singularities that dominate the line shape for small excitation energy $(\omega -\omega_L^{\tau} (k))$ 
near the lower thresholds $\omega_L^{\tau} (k)$ of the longitudinal $(\tau = l)$ and transversal $(\tau = t)$ dynamical structure 
factors $S^{zz} (k,\omega)$ and $S^{xx} (k,\omega)$, respectively, Eq. (\ref{SDSF}), are within the PDT
determined by class (ii) excitations generated by specific leading-order processes in the BA distributions. 
(Class (ii) excitations are $\vert S^z\vert =S$ excited states \cite{Muller}.) At 
the present isotropic $\Delta =1$ point, higher-order class (ii) excitations generated by additional 
higher-order processes in the BA distributions also contribute to the dynamical structure factors but in the TL 
they change neither the momentum dependent exponents nor the corresponding singularities sharpness.
Moreover, class (ii) excitations described by complex BA rapidities are gapped for $h> 0$ and 
except for very small $h$ have nearly vanishing spectral weight. 
For instance, at $m=0.5$ their contributions correspond to a relative intensity of about $3\times 10^{-7}$ for anisotropy $\Delta =0.3$ and 
$4\times 10^{-7}$ for $\Delta =0.7$ \cite{Caux-05}. Even for  $\Delta =1$,  the corresponding estimated relative intensity is not
larger than $10^{-6}$. Our study focuses mainly on the $m>0.15$ range for which their contribution is negligible.

The lower threshold singularities of the dynamical structure factors, 
Eq. (\ref{SDSF}), are for $h> 0$ determined by excitations whose BA rapidities $\lambda_j$ are real. 
There is a one-to-one correspondence, $\lambda_j =\lambda (q_j)$, to the BA quantum numbers
$q_j = {2\pi\over L} I_j$ where $j = 1,\cdots,L-M$ and $I_j$ are successive half-odd integers for $L-M$ 
even and integers for $L-M$ odd. Out of the set of $j = 1,\cdots,L-M$ quantum
numbers $\{q_j\}$, a subset of $\alpha = 1,\cdots,M$ numbers $\{q_{\alpha}\}$ are
occupied. These refer to the "BA band particles" whereas the remaining $L-2M=L\,m$ values $q_j$ 
correspond to the "BA band holes", which here are simply called particles and holes, respectively.
The momentum of these states is $k = M\,\pi + \sum_{\alpha} q_{\alpha}$,
so that the numbers $\{q_j\}$ indeed play the role of BA band momentum values.

We consider the TL, $L\rightarrow\infty$, within which the set of $j = 1,\cdots,L-M$ 
momentum values $\{q_j\}$ (such that $q_{j+1}-q_j = {2\pi\over L}$)
may be replaced by a continuum momentum variable, $q \in [-k_{F\uparrow},k_{F\uparrow}]$,
and the rapidities $\lambda_j =\lambda (q_j)$ by a rapidity function, $\lambda =\lambda (q)\in [-\infty,\infty]$,
with $\lambda (\pm k_{F\uparrow})=\pm\infty$. Except for ${\cal{O}} (1/L)$ corrections, 
one has that $k_{F\uparrow}={\pi\over 2} (1+m)$
and $k_{F\downarrow}={\pi\over 2}  (1-m)$ with $q \in [-k_{F\downarrow},k_{F\downarrow}]$
corresponding to the occupied momentum values of $m\in [0,1]$ ground states. Hence for $m> 0$ 
such states are populated by holes for $\vert q\vert\in [k_{F\downarrow},k_{F\uparrow}]$. 
The BA band is full for the $m=0$ ground state. (Its excited states holes are usually identified with spin-$1/2$
spinons \cite{Caux-11,Rodrigo-08,Essler-14}, whereas those of $m>0$ states are called here holes \cite{note}.) 
We denote by $\lambda_0 (q)$ the ground state rapidity function such that $\lambda_0 (q)=-\lambda_0 (-q)$,
$\lambda_0 (k_{F\downarrow}) = B$, and $\lambda_0 (k_{F\uparrow}) = \infty$ where $B=\infty$ for $m\rightarrow 0$ and
$B=0$ for $m\rightarrow 1$. It can be defined in terms of its 
inverse function, $q = \int_0^{\lambda_0 (q)}d\lambda \,2\pi\sigma (\lambda)$,
where the usual BA distribution $2\pi\sigma (\lambda)$ is the solution of the integral equation $2\pi\sigma (\lambda) = {4\over 1 + (2\lambda)^2} 
- \frac{1}{\pi} \int_{-B}^{B}d\lambda^{\prime}\,{2\pi\sigma (\lambda^{\prime})\over 1 + \left(\lambda -
\lambda^{\prime}\right)^2}$. 

The spectra, $\omega^{\tau} (k) = \omega^{\tau} (-k)$, of the excited states that for $h> 0$ and spin densities $m\in ]0,1]$ control the
leading order contributions to the dynamical structure factors involve the BA band dispersion 
$\varepsilon (q) = \varepsilon^0 (q) - \varepsilon^0 (k_{F\downarrow})$
where $q \in [-k_{F\uparrow},k_{F\uparrow}]$, $\varepsilon^0 (q) = {\bar{\varepsilon}}^{\,0} (\lambda_0 (q))$, and,
\begin{eqnarray}
{\bar{\varepsilon}}^{\,0} (\lambda) & = & - {2J\over 1 + (2\lambda)^2} 
+ 4J\int_{-B}^{B}d\lambda^{\prime}\,{\lambda^{\prime}\over 1 + (2\lambda^{\prime})^2}\,\bar{\Phi } (\lambda^{\prime},\lambda)
\nonumber \\
\bar{\Phi } (\lambda,\lambda^{\prime}) & = & {\arctan (\lambda -\lambda^{\prime})\over\pi}
- \int_{-B}^{B}{d\lambda''\over\pi}{\bar{\Phi } (\lambda'',\lambda^{\prime})
\over 1 + (\lambda -\lambda'')^2} \, .
\label{bar-phase-shift}
\end{eqnarray}
The latter integral equation defines the rapidity-dependent dressed phase shift $2\pi\bar{\Phi } (\lambda,\lambda')$ 
in units of $2\pi$. The group velocity reads $v (q) = \partial\varepsilon (q)/\partial t = \partial\varepsilon^0 (q)/\partial t$, with
the Fermi velocity $v = v (k_{F\downarrow})$ playing an important role in the low-energy physics.
The dispersion  $\varepsilon^0 (q)$ controls the spin density curve, 
$h (m) = - \varepsilon^0 (k_{F\downarrow})/(2\mu_B)$ with $m = 1 - 2k_{F\downarrow}$. The range $m \in ]0,1]$ 
refers to $h\in ]0,h_c]$ where $h_c = J/\mu_B$ is the critical field at which fully polarized ferromagnetism 
is achieved.

The processes that generate the excited states that control the
lower threshold singularities correspond to specific values of the right ($\iota=1$) and left ($\iota =-1$) Fermi points particle 
number deviations $\delta N_{F\iota}$. It is more convenient though to use the corresponding number deviations
$\delta N_F = \sum_{\iota=\pm 1}\delta N_{F\iota}$ and $\delta J_F = {1\over 2}\sum_{\iota=\pm 1}(\iota)\delta N_{F\iota}$.
Within the PDT, the exponents that control the line shape in the vicinity of the theory branch lines that
coincide with $\tau =l,t$ lower thresholds involve the functional,
\begin{equation}
2\Delta_{\iota}^{\tau} (q) = \left(\iota{\delta N_F\over 2\xi_0}
+ \xi_0\,\delta J^F + C\,\Phi(\iota k_{F\downarrow},q)\right)^2 \, .
\label{functional}
\end{equation}
Here $\iota =\pm 1$, the momentum $q$ is that of a particle created ($C=1$) or annihilated
($C=-1$) under the transitions to the excited states, 
$\Phi (q,q') = \bar{\Phi } (\lambda_0 (q),\lambda_0 (q'))$ is a momentum dependent dressed
phase shift, $\bar{\Phi } (\lambda,\lambda^{\prime})$ obeys Eq. (\ref{bar-phase-shift}), and the related 
parameter, $\xi_0 = 1+ \Phi(k_{F\downarrow},k_{F\downarrow}) - \Phi(k_{F\downarrow},-k_{F\downarrow})$, 
increases from $\xi_0 = 1/\sqrt{2}$ for $m\rightarrow 0$ to $\xi_0 = 1$ as $m\rightarrow 1$. 
Indeed, for $m\rightarrow 0$ the dressed phase shift $\Phi (q,q')$ has the limiting value
$\Phi (\iota\,k_{F\downarrow},q) = \iota/(2\sqrt{2})$ for $q\neq \iota\,k_{F\downarrow}$ 
whereas $\Phi(\iota\,k_{F\downarrow},\iota\,k_{F\downarrow}) = \iota (3 - 2\sqrt{2})/(2\sqrt{2})$.

The longitudinal spectrum, $\omega^{l} (k) = \omega^{l} (-k)$, refers to excited states that are generated by one
particle-hole processes and conserve $M$,
\begin{equation}
\omega^l (k) = - \varepsilon (q_1) + \varepsilon (q_2)  \, ; \hspace{0.25cm} k = q_2 - q_1 \, .
\label{dkEdP}
\end{equation}
On the other hand, the spectrum $\omega^t (k)$ of $S^{xx} (k,\omega) = {1\over 4}[S^{+-} (k,\omega)+S^{-+} (k,\omega)]$
is here expressed as the superposition of the spectra (i) $\omega^{+-} (k)$ and (ii) $\omega^{-+} (k)$.
The spectra $\omega^{\pm\mp} (k)$ refer to states that are generated by a zero-energy 
$\delta M=\delta N_{F\iota}=\pm 1$ process at a $\iota$ Fermi point along with an overall
BA band shift $\delta q_j=\mp \iota\pi/L$, which renders it a net zero-momentum process, plus one (i) 
particle-hole and (ii) two-hole process,
\begin{eqnarray}
\omega^{+-} (k) & = & \varepsilon (q_2) - \varepsilon (q_1) \, ; \hspace{0.25cm} k = \pi + q_2 - q_1 \, ,
\nonumber \\
\omega^{-+} (k) & = & - \varepsilon (q_1) - \varepsilon (q_2) \, ; \hspace{0.25cm} k = \pi  - q_1 - q_2 \, .
\label{dkEdPxx}
\end{eqnarray}
All above spectra refer to spin densities $m\in ]0,1]$ and momentum $k \in [0,\pi]$. In Eqs. (\ref{dkEdP})
and (\ref{dkEdPxx}), $q_1 \in [-k_{F\downarrow},k_{F\downarrow}]$ and $q_2 \in [k_{F\downarrow},k_{F\uparrow}]$ 
for the longitudinal spectrum, $q_2 \in [-k_{F\uparrow},-k_{F\downarrow}]$ for 
the $+-$ spectrum, and $q_2 \in [-k_{F\downarrow},k_{F\downarrow}]$ for the $-+$ spectrum.

In the $m\rightarrow 0$ limit, the spectra $\omega^l (k)$, Eq. (\ref{dkEdP}), and $\omega^{+-} (k)$, 
Eq. (\ref{dkEdPxx}), reduce to their lower thresholds as $h\rightarrow 0$, which in that limit becomes the lower 
threshold of both $\omega^{-+} (k)$, Eq. (\ref{dkEdPxx}), and the class (ii) two-hole excitations described by complex rapidities 
whose gap vanishes as $h\rightarrow 0$. At $h=0$ the spectrum $\omega^{-+} (k)$ is also that
of the $S^z =0$ and $S=1$ two-hole excitations of class (i), which due to a selection rule \cite{Muller} 
do not contribute to the dynamical structure factors at $h> 0$. Hence upon smoothly turning off $h$ 
there is a large weight transfer from $\vert S^z\vert =S$ class (ii) excitations for $h\rightarrow 0$ to 
degenerate $S^z =0$ and $S=1$ class (i) two-hole states at $h=0$. 

A {\it particle} (and {\it hole}) {\it branch line} is for the PDT generated by excitations where
one particle is created (and annihilated) outside the $q=\pm k_{F\downarrow}$ Fermi points
and all remaining processes occur at such points. For both spin densities $m\rightarrow 0$ and $m>m_*\approx 0.15$, 
the lower threshold of $\omega^l (k)$ (and $\omega^t (k)$) coincides with a hole branch line for $k \in [0,2k_{F\downarrow}]$
(and $k \in [\pi - 2k_{F\downarrow},\pi]$) and with a particle branch line for $k \in [2k_{F\downarrow},\pi]$
(and $k \in [0,\pi -2k_{F\downarrow}]$). For $0<m<m_*\approx 0.15$, the branch-line spectrum $\omega^l (k)$ 
(and $\omega^t (k)$) does not coincide with the hole branch line for a small momentum width 
near $k=0$ (and $k=\pi$.) For simplicity, we consider mostly spin densities $m\rightarrow 0$ and 
$m>m_*\approx 0.15$ for which $\omega_{L}^{\tau} (k)$ coincides with branch lines and the PDT gives 
the exact momentum and spin density dependence of the exponents that control the line shape in its vicinity. 
(Even for $0<m<m_*\approx 0.15$ they are a good approximation.) 

Interestingly, the use of the PDT reveals that the lower threshold singularities of $S^{xx} (k,\omega)$
are those of $S^{+-} (k,\omega)$ near the particle branch line and of $S^{-+} (k,\omega)$ near the hole branch line.
Accounting for $\varepsilon (\pm k_{F\downarrow}) =0$,
the longitudinal $S^{zz} (k,\omega)$ and transversal $S^{xx} (k,\omega)$ hole branch lines spectra are given by
\begin{eqnarray}
\omega_h^{\tau} (k) & = & - \varepsilon (q) \, , \hspace{0.25cm} \tau = l,t \, ,
\nonumber \\
k & = & k_{F\downarrow} - q \in [0,2k_{F\downarrow}] \, , \hspace{0.25cm} \tau = l \, ,
\nonumber \\
k & = & \pi -k_{F\downarrow} - q \in [\pi - 2k_{F\downarrow},\pi] \, , \hspace{0.25cm} \tau = t \, ,
\label{dEdP-HBL}
\end{eqnarray}
where $q \in [-k_{F\downarrow},k_{F\downarrow}]$. The corresponding particle branch lines spectra are given by,
\begin{eqnarray}
\omega_p^{\tau} (k) & = & \varepsilon (q) \, , \hspace{0.25cm} \tau = l,t \, ,
\nonumber \\
k & = & k_{F\downarrow} + q \in [2k_{F\downarrow},\pi] \, , \hspace{0.25cm} \tau = l \, ,
\nonumber \\
k & = & \pi - k_{F\downarrow} + q \in [0,\pi - 2k_{F\downarrow}] \, , \hspace{0.25cm} \tau = t \, ,
\label{dEdP-PBL}
\end{eqnarray}
with $q \in [k_{F\downarrow},k_{F\uparrow}]$ and $q \in [-k_{F\uparrow},-k_{F\downarrow}]$ for the 
$l$ and $t$ particle branch lines, respectively.

For the $C=1$ particle and $C=-1$ hole $l$ branch lines the Fermi points number deviations to be 
used in Eq. (\ref{functional}) are $\delta N_F = -C$ and $\delta J_F = {1\over 2}$.
For the $t$ branch lines such deviations are given by $\delta N_F = 0$ and $\delta J_F = {1\over 2}$.
One then finds,
\begin{eqnarray}
2\Delta_{\iota}^{l} (q)  & = & \left({\xi_0^2 - \iota\,C\over 2\xi_0} + C\,\Phi(\iota k_{F\downarrow},q)\right)^2 \, ,
\nonumber \\
2\Delta_{\iota}^{t} (q) & = & \left({\xi_0\over 2} + C\,\Phi(\iota k_{F\downarrow},q)\right)^2 \, ,
\label{deltas-lt}
\end{eqnarray}
where $q \in [-k_{F\downarrow},k_{F\downarrow}]$ for $C=-1$ and $\tau = l,t$, $q \in [k_{F\downarrow},k_{F\uparrow}]$ 
for $C=1$ and $\tau = l$, and $q \in [-k_{F\uparrow},-k_{F\downarrow}]$ for $C=1$ and $\tau = t$.
\begin{figure}[hbt]
\begin{center}
\centerline{\includegraphics[width=6.50cm,angle=-90]{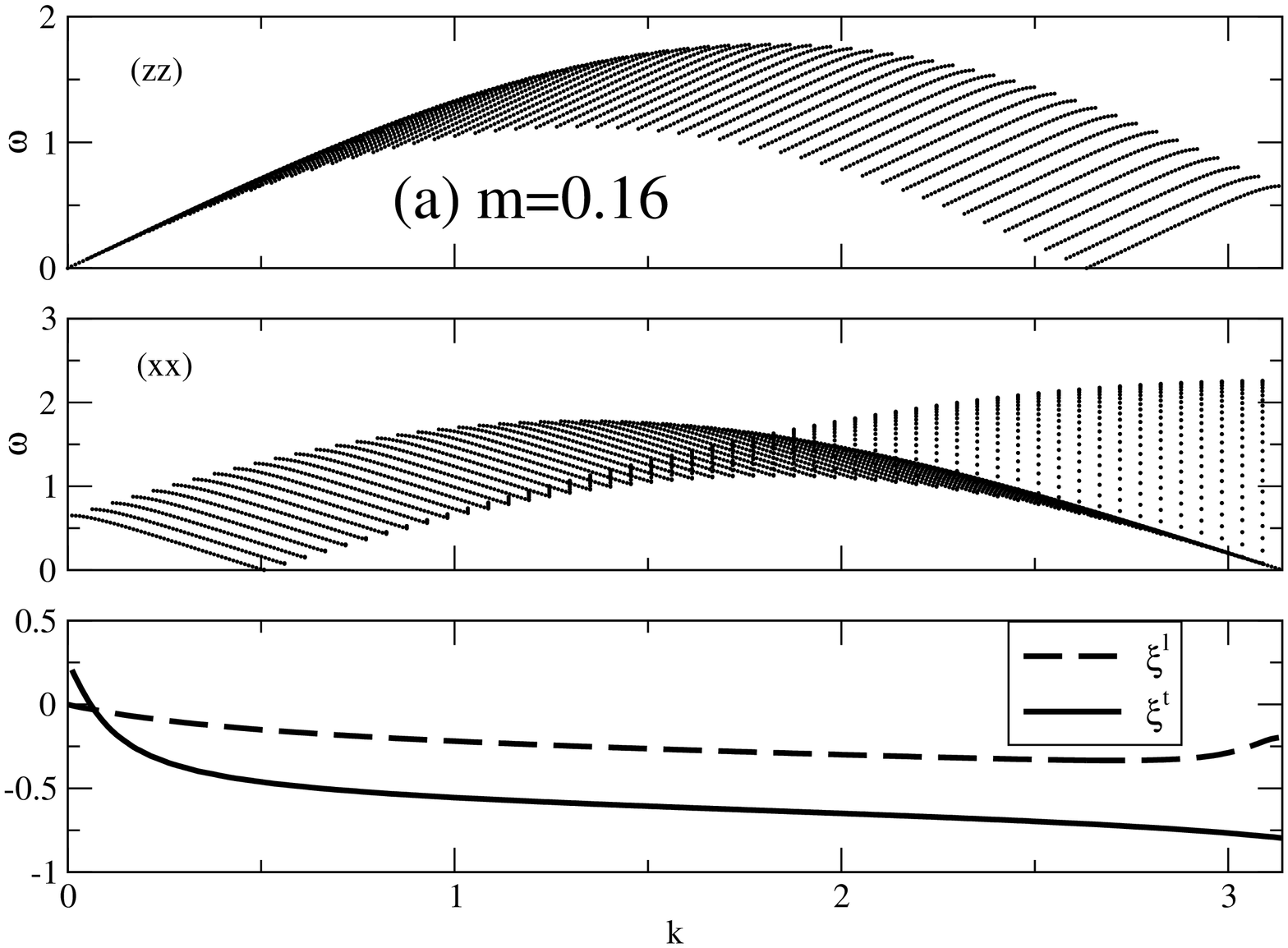}}
\centerline{\includegraphics[width=6.50cm,angle=-90]{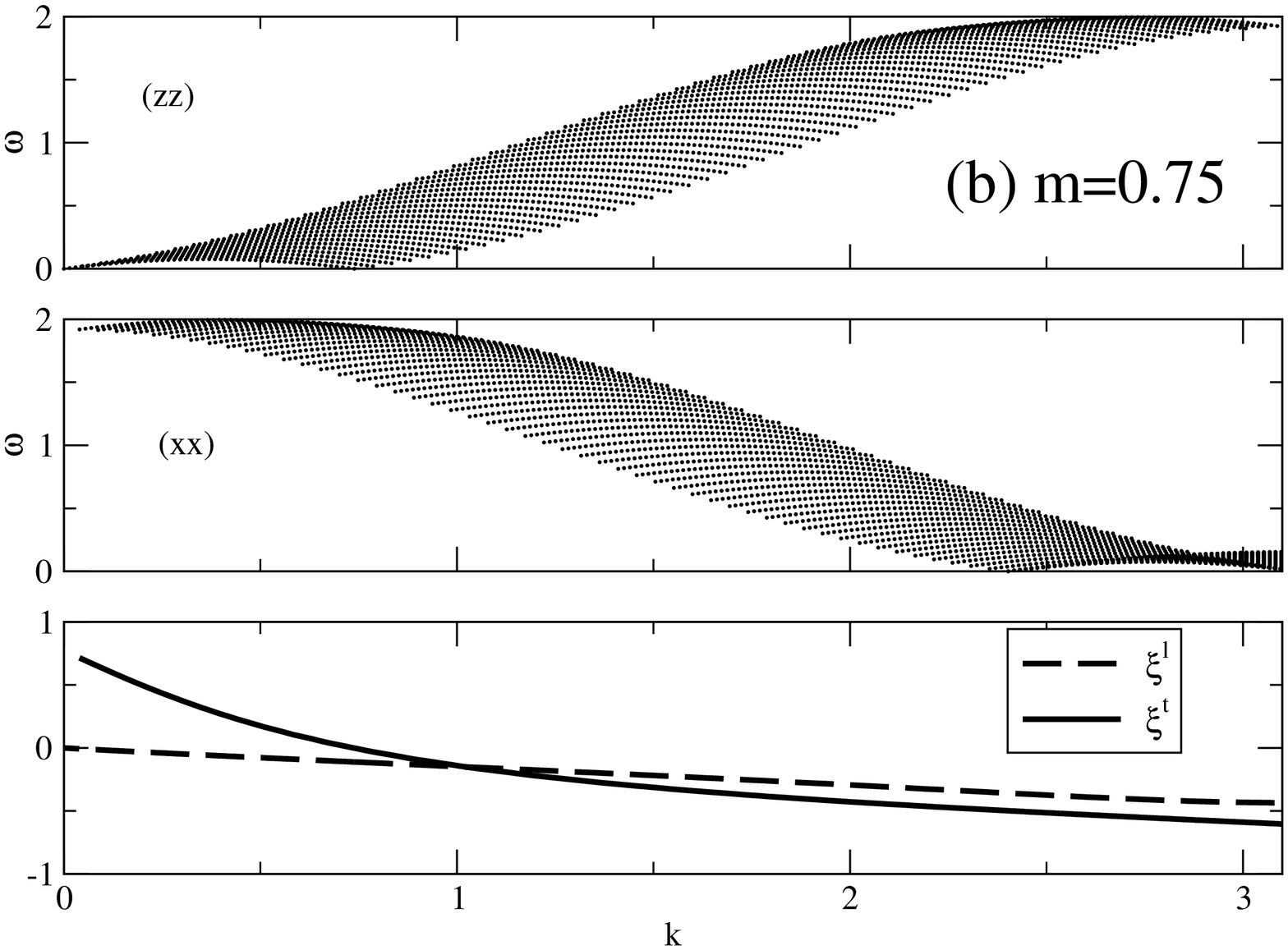}}
\caption{Two upper panels of each sub-figure (a) and (b): 
The spectra $\omega^l (k)$ and $\omega^t (k)$ for (a) $m=0.16$ and (b) $m=0.75$.
Lower panels: The exponents $\xi^{l} (k)$ and $\xi^{t} (k)$, Eq. (\ref{DSF-BL}), that control the singularities
in the vicinity of the lower thresholds of the spectra plotted here as a function of $k\in [0,\pi]$.}
\label{fig1}
\end{center}
\end{figure}

For small positive values $(\omega - \omega_{L}^{\tau} (k))$ in the vicinity of $\omega_{L}^{\tau} (k)>0$, the 
dynamical structure factors are according to the PDT of the form,
\begin{eqnarray}
S^{\alpha\alpha} (k,\omega) & = & C_L^{\tau}\,(\omega - \omega_{L}^{\tau} (k))^{\xi^{\tau} (k)} \, , \hspace{0.25cm} k \in [0,\pi] \, ,
\nonumber \\
\xi^{\tau} (k) & = & -1 + \sum_{\iota=\pm 1}2\Delta_{\iota}^{\tau} (q) \, .
\label{DSF-BL}
\end{eqnarray}
Here $\alpha =z$ for $\tau =l$, $\alpha =x$ for $\tau =t$, $C_L^{\tau}$ is a $k$ and $\omega$ independent constant, 
and the $k$ values that correspond to $q$ in $2\Delta_{\iota}^{\tau} (q)$ are those of Eqs. (\ref{dEdP-HBL}) and (\ref{dEdP-PBL}). 

The exponent $\xi^{\tau} (k)$ given in Eq. (\ref{DSF-BL}) does not apply near the $\omega=0$ lower threshold soft modes
such as $(k_0^{\tau},0)$ where $k_0^{l}=2k_{F\downarrow}$ and $k_0^{t}=\pi -2k_{F\downarrow}$ in the $(k,\omega)$-plane.  
In this case the PDT reaches the same results as conformal field theory (CFT).
Indeed, near them the functionals, Eq. (\ref{functional}), become the conformal dimensions of
the $\iota =\pm 1$ fields \cite{LE}, $2\Delta_{\iota}^{\tau} = (\iota\delta N_F/(2\xi_0) + \xi_0\,\delta J^F)^2$. Here
$\delta N_F = 0$, $\delta J_F = 1$ for $\tau =l$ and $\delta N_F = \pm 1$, $\delta J_F = -1$ for $\tau =t$.
While in this Letter we are mostly interested in the finite-energy behavior
of $S^{\alpha\alpha} (k,\omega)$,  for completeness we provide its general form near $(k_0^{\tau},0)$, which can
be obtained from CFT. Specifically, one finds $S^{\alpha\alpha} (k,\omega) =  C_{\pm 1}^{\tau}\,(\omega)^{\zeta_{\pm 1}^{\tau}}$
near the branch lines for small $\omega \approx \pm v\,(k-k_0^{\tau})$ and $S^{\alpha\alpha} (k,\omega) =  C_{0}^{\tau}\,(\omega)^{\zeta_0^{\tau}}$ 
away from such lines for small $\omega \neq \pm v\,(k-k_0^{\tau})$. Here $C_{\pm 1}^{\tau}$ and $C_0^{\tau}$ are 
$k$ and $\omega$ independent constants and the exponents read $\zeta_{\pm 1}^{\tau} = -1+ 2\Delta_{\pm}^{\tau}$
and $\zeta_0^{\tau} = -2+\sum_{\iota =\pm 1}2\Delta_{\iota}^{\tau}$. 

In the present $L\rightarrow\infty$ limit, the $h>0$ higher-order particle-hole excitations do not change the
exponent and corresponding sharpness of the singularities, Eq. (\ref{DSF-BL}). They may however change slightly 
the values of the $k$ and $\omega$ independent constants $C_L^{\tau}$. For finite systems
they give rise to a small tail for small negative values $(\omega - \omega_{L}^{\tau} (k))$. 
Such finite-$L$ effects are seen in Fig. 1 of Ref. \cite{Caux-05} in $S^{zz} (k,\omega)$ for anisotropy 
$\Delta =0.7$ and momentum values $k=3\pi/4$ and $k=\pi/2$. Due to the finiteness of 
$\log L/L$, the peaks appear sharper if one does not account for the contributions from the two 
particle-hole excitations. Our results refer to $L\rightarrow\infty$ so that the latter contributions do 
not affect the sharpness of the peaks, Eq. (\ref{DSF-BL}), since $\log L/L\rightarrow 0$ as $L\rightarrow\infty$.

The spectrum $\omega^l (k)$, Eq. (\ref{dkEdP}), and the spectrum $\omega^t (k)$ that results
from combination of the spectra $\omega^{+-} (k)$ and $\omega^{-+} (k)$, Eq. (\ref{dkEdPxx}),
along with the corresponding exponents $\xi^{l} (k)$ and $\xi^{t} (k)$ given in Eq. (\ref{DSF-BL}) are
plotted in Fig. \ref{fig1} for spin densities (a) $m=0.16$ and (b) $m=0.75$. The exponent $\xi^{l} (k)$ is negative 
for $k>0$ at any $m$ value whereas $\xi^{t} (k)$ is negative for a $m$-dependent range $k\in [k_t,\pi]$.
Here $k_t$ increases from $k_t=0$ for $m\rightarrow 0$ to $k_t = -2\arctan\left({1\over 2}\tan(\pi/\sqrt{2})\right)
\approx 0.37\,\pi$ for $m\rightarrow 1$. For the $k$ ranges for which $\xi^{\tau} (k)<0$, there are
lower threshold singularity cusps in $S^{\alpha\alpha} (k,\omega)$, Eq. (\ref{DSF-BL}). 

In the $m\rightarrow 0$ limit both the $\tau = l,t$ lower thresholds $\omega_{L}^{\tau} (k)$ coincide with the hole branch line
for all $k$ values. In that limit the corresponding exponents $\xi^{\tau} (k)$ are given by $\xi^{\tau} (k)=-1/2$ for 
all $k$ values and the lower thresholds $\omega_{L}^{\tau} (k)$  
coincide with that of the $m=0$ two-hole spectrum. Consistently, $\xi^{\tau} (k)=-1/2$ is also the value of the known
exponent that controls the line shape in the vicinity of the lower threshold of the
latter spectrum \cite{Bougourzi-97,Imambekov-09,Caux-11}.

In the opposite limit, $m\rightarrow 1$, the lower thresholds $\omega_{L}^{\tau} (k)$ coincide with the particle branch line 
for all $k$ values and the $\tau = l,t$ exponents read $\xi^l (k) = 2\Phi (0,k)[1+\Phi (0,k)]$ and
($\xi^t (k) = -1/2 + 2\Phi (0,k-\pi)[1+\Phi (0,k-\pi)]$ where
the phase shift reads $\Phi (0,q) = - {1\over\pi}\arctan\left({1\over 2}\tan \left({q\over 2}\right)\right)$. 
In this limit, $\xi^l (k) = 0$ (and $\xi^t (k) = 1$) for $k\rightarrow 0$ decreases to $\xi^l (k) = - 1/2$ (and $\xi^t (k) = - 1/2$) 
for $k\rightarrow \pi$. The corresponding $m\rightarrow 1$ behaviors refer to a small but finite $M/L$ value. 
For both zero and finite $M$ values reached as $h\rightarrow h_c$, the longitudinal spin dynamical structure factor 
vanishes in the TL \cite{Muller}, the superimposed dynamical structure factor being for $h\rightarrow h_c$ 
dominated by $S^{xx} (k,\omega)$. At $h=h_c$ Eq. (\ref{DSF-BL}) is not valid, being replaced by
the $\delta$-function like distribution,
$S^{xx} (k,\omega) = {\pi\over 2} \delta \left(\omega - J (1+\cos k)\right)$ for $k  \in  [0,\pi]$.

The structure form factors $S^{zz} (k,\omega)$ and $S^{xx} (k,\omega)$ may be investigated separately in
$h>0$ experiments on spin-chain compounds by using a carefully oriented crystal. If the crystal is misoriented, or if a micro crystalline sample
is used, the $S^{zz} (k,\omega)$ and $S^{xx} (k,\omega)$ spectral features should appear superimposed.
Such superimposition changes the excitations lower thresholds and leads to the broadening of the singularities,
Eq. (\ref{DSF-BL}). However, this does not occur at $h=0$, since $S^{zz} (k,\omega)=S^{xx} (k,\omega)$.

These two different situations are clearly seen in the magnetic scattering intensity measured 
at zero- and finite-field inelastic neutron scattering experiments of Ref. \cite{Stone-03}, respectively,  
on Cu(C$_4$H$_4$N$_2$)(NO$_3$)$_2$. In Figs. 2 (a)-(c) of Ref. \cite{Stone-03} the theoretically predicted sharp cusps at zero-field, 
$S^{zz} (k,\omega)=S^{xx} (k,\omega)=C_L\,(\omega - \omega_{L} (k))^{-1/2}$, are clearly seen
at different $k$ values. On the other hand, the $S^{zz} (k,\omega)\neq S^{xx} (k,\omega)$ spectral features appear 
superimposed in the finite-field Figs. 2 (d)-(f) of that reference, so that only at $k\approx \pi$ is the theoretically predicted 
sharp cusp clearly visible. 

In summary, we have obtained the exact momentum dependence of the exponents that in the TL
control the line shape singularities in the vicinity of the lower thresholds of the longitudinal and
transverse dynamical spin structure factors using the PDT \cite{V-1,LE}, in the fundamental case 
of the spin-$1/2$ $XXX$ Heisenberg chain in a field $h> 0$. We suggest that more demanding $h> 0$ experiments with a 
carefully oriented crystal are carried out on Cu(C$_4$H$_4$N$_2$)(NO$_3$)$_2$ and other spin-chain compounds,
thus yielding separately $S^{zz} (k,\omega)$ and $S^{xx} (k,\omega)$ whose magnetic scattering intensities are expected to 
display the singularity cusps theoretically studied in this Letter.

We thank A. H. Castro Neto, H. Q. Lin, and T. Prosen for discussions. J. M. P. C. and P. D. S.
thank the support by the Beijing CSRC and the FEDER through the 
COMPETE Program and the Portuguese FCT in the framework of the Strategic Projects PEST-C/FIS/UI0607/2013 and 
PEST-OE/FIS/UI0091/201, respectively.


\begin{thebibliography}{99}
\bibitem{Taka-AN}
	M. Takahashi and M. Suzuki, Prog. Theor. Phys. {\bf 48}, 2187 (1972).
\bibitem{Taka}
	M. Takahashi, Prog. Theor. Phys. {\bf 46}, 401 (1971).
\bibitem{Muller}		
	G. M\"uller, H. Thomas, H. Beck, and J. C. Bonner,
	Phys. Rev. B {\bf 24}, 1429 (1981).
\bibitem{SPA-11}
	J. Sirker, R. G. Pereira, and I. Affleck,
	Phys. Rev. B {\bf 83}, 035115 (2011).
\bibitem{Stone-03}
	M. B. Stone, D. H. Reich, C. Broholm, K. Lefmann, C. Rischel, C. P. Landee, and M. M. Turnbull,
	Phys. Rev. Lett. {\bf 91}, 037205 (2013).
\bibitem{Thurber-01}
	K. R. Thurber, A. W. Hunt,T. Imai, and F. C. Chou,
	Phys. Rev. Lett. {\bf 87}, 247202 (2001).
\bibitem{Motoyama-96}
	N. Motoyama, H. Eisaki, and S. Uchida,
	Phys. Rev. Lett. {\bf 76}, 3212 (1996).
\bibitem{Bougourzi-97} 
	M. Karbach, G. M\"uller, A. H. Bougourzi, A. Fledderjohann,
	and K. H. M\"utter, Phys. Rev. B {\bf 55}, 12510 (1997);
	A. Abada, A. H. Bougourzi, and B. Si-Lakhal, 
	Nucl. Phys. B {\bf 497}, 733 (1997); A. H. Bougourzi, M. Couture, and M. Kacir,
	Phys. Rev. B {\bf 54}, R12669 (1996).
\bibitem{Caux-06}		
	J. - S. Caux and R. Hagemans,
	J. Stat. Mech. P12013 (2006).
\bibitem{Imambekov-09}
	A. Imambekov and L. I. Glazman, 
	Phys. Rev. Lett. {\bf 102}, 126405 (2009). 
\bibitem{Caux-11}		
	J. - S. Caux, H. Konno, M. Sorrell, and R. Weston,
	Phys. Rev. Lett. {\bf 106}, 217203 (2011).
\bibitem{Lefmann-96}
	K. Lefmann and C. Rischel,
	Phys. Rev. B {\bf 54}, 6340 (1996).
\bibitem{Karbach-00}	
	M. Karbach and G. M\"uller, Phys. Rev. B {\bf 62}, 14871 (2000).
\bibitem{Karbach-02}
	M. Karbach, D. Biegel, and G. M\"uller,
	Phys. Rev. B {\bf 66}, 054405 (2002).	
\bibitem{Biegel-02}
	D. Biegel, M. Karbach, and G. M\"uller,
	Europhys. Lett. {\bf 59}, 882 (2002).
\bibitem{Kitanine-99}
	N. Kitanine, J. M. Maillet, and V. Tetras,
	Nucl. Phys. B {\bf 554}, 647 (1999). 
\bibitem{Caux-05}
	J.-S. Caux and J. M. Maillet,
	Phys. Rev. Lett. {\bf 95}, 077201 (2005). 
\bibitem{Rodrigo-08}
	R. G. Pereira, S. R. White, and I. Affleck, Phys. Rev. Lett. {\bf 100}, 027206 (2008).	
\bibitem{V-1}
        J. M. P. Carmelo, K. Penc, D. Bozi, Nucl. Phys. B {\bf 725}, 421  (2005) ; {\bf 737}, 351, Erratum (2006);
        J. M. P. Carmelo, K. Penc, P. D. Sacramento, M. Sing, and R. Claessen, J. Phys.: Cond. Mat. {\bf 18}, 5191 (2006);
        J. M. P. Carmelo, D. Bozi, and K. Penc, J. Phys.: Cond. Mat. {\bf 20}, 415103 (2008).
\bibitem{LE}
        J. M. P. Carmelo, L. M. Martelo, K. Penc, Nucl. Phys. B {\bf 737}, 237 (2006).
\bibitem{Hubbard-94}
	F. H. L. Essler, H. Frahm, F. G\"ohmann, A. Kl\"umper, V. E. Korepin, in
	{\em The one-dimensional Hubbard model} (Cambridge University Press, Cambridge, UK, 2005);
	J. M. P. Carmelo, A. H. Castro Neto, and D. K. Campbell,
	Phys. Rev. B {\bf 50}, 3667 (1994); J. M. P. Carmelo, A. H. Castro Neto, and D. K. Campbell,
	Phys. Rev. B {\bf 50}, 3683 (1994).
\bibitem{Essler-14}        
        L. Seabra, F. H. L. Essler, F. Pollmann, I. Schneider, and T. Veness,
        Phys. Rev. B {\bf 90}, 245127 (2014).
\bibitem{Glazman-12}  
	A. Imambekov, T. L. Schmidt, and L. I. Glazman,
	Rev. Mod. Phys. {\bf 84}, 1253 (2012).
\bibitem{DSF-n1}            
       	R. G. Pereira, K. Penc, S. R. White, P. D. Sacramento J. M. P. Carmelo,
	Phys. Rev. B {\bf 85}, 165132 (2012).  
\bibitem{note}  	
	Elsewhere it will be shown that the holes of $m> 0$ ground states and their excited states 
	have scattering properties different from those of spin-$1/2$ objects, such as the $m=0$ spinons. 	
\end{thebibliography}
\end{document}